\input f.sty

\input epsf.sty
\magnification1020

\rightline\timestamp
\rightline{FTUAM 04-01}
\rightline{hep-ph/XX}
\bigskip
\hrule height .3mm
\vskip.6cm
\centerline{{\bigfib Scattering amplitudes at  multi TeV energies}\footnote{*}{
Dedicated to Prof.~Yuri Simonov, one of the first physicists to establish the connection 
between Regge theory and QCD.}}
\medskip
\centerrule{.7cm}
\vskip1cm

\setbox9=\vbox{\hsize65mm {\noindent\medfib F. J. 
Yndur\'ain} 
\vskip .1cm
\noindent{\addressfont Departamento de F\'{\i}sica Te\'orica, C-XI\hb
 Universidad Aut\'onoma de Madrid,\hb
 Canto Blanco,\hb
E-28049, Madrid, Spain.}\hb}
\smallskip
\centerline{\box9}
\bigskip
\setbox0=\vbox{\abstracttype{Abstract} We show that a generalized Regge 
 behaviour,
$$\imag F(s,t)\simeq
\phiv(t)(\log s/\hat{s})^{\nu(t)}(s/\hat{s})^{\alpha_P(t)},\quad{\rm for}\; |t|<|t_0|,\;
s\to\infty$$
where $\phiv(t)\simeq \ee^{bt}$, $\alpha_P(t)\simeq \alpha_P(0)+\alpha'_P(0)t$, 
and $t_0$ is the first zero of $\alpha_P(t)$,  $\alpha_P(t_0)=0$,
 implies that the corresponding cross section is bounded by
$$\sigma_{\rm tot}(s)<({\rm Const.})\times\log s/\hat{s}.$$
This growth, however, is not sufficient to fit the experimental cross sections. 
If, instead of this, we assume  saturation of 
the improved Froissart bound, i.e., a behaviour
$$\imag F(s,0)\simeq A(s/\hat{s})\log^2{{s}\over{s_1\log^{7/2} s/s_2}},
$$ a good fit 
is obtained to $\pi\pi$, $\pi N$, $KN$ and $NN$ cross sections from 
c.m. kinetic energy $E_{\rm kin}\simeq1\,$~GeV to 30 TeV 
(producing a cross section of $108\pm6$ mb at LHC energy). 
This suggests that the Regge-type behaviour only holds 
for values of the momentum transfer very near zero.
}
\centerline{\box0}

\brochureendcover{Typeset with \physmatex}
\brochureb{\smallsc f. j.  yndur\'ain}{\smallsc 
scattering amplitudes at  multi tev energies}{1}

\booksection{1. Introduction}

\noindent
It has been known for a long time\ref{1,2} that the analyticity and unitarity properties for 
scattering amplitudes that follow from local field theory (or, more generally, 
in a theory where the {\sl observables} are local\ref{2}) imply the Froissart 
bound for total hadronic  cross sections, $A+B\to{\rm all}$,
$$\sigma_{AB}(s)\lsim_{s\to\infty}C\log^2s/s_0.
\equn{(1.1a)}$$ 
In particular, for $\pi\pi$ scattering, we can calculate the constants $C$, $s_0$ 
and the corrections to (1.1a) for finite energies, the last in terms of the 
pion mass, $m_\pi$, and the D wave scattering lengths.\ref{3}
In fact, (1.1a) can be somewhat improved in the sense that one can show\ref{4} that $s_0$ 
must grow as a \ffrac{7}{2} power\fnote{Note 
that in ref.~4 the power is wrongly given as 7
 instead of the correct value, \ffrac{7}{2}.}  of a logarithm of the energy, so 
 one has the bound
$$\sigma_{AB}(s)\lsim_{s\to\infty}C\log^2\dfrac{s}{s_1\log^{\frac{7}{2}}s/s_2},
\equn{(1.1b)}$$
and $s_1$, $s_2$ are now constants.

Experimental cross sections seem to be close to the bounds~(1.1)
in the sense that they exhibit  growth\ref{5} at very high ($s^{1/2}>10\,\gev$) 
energies. But it has up till 
now not been possible to calculate the {\sl behaviour} 
(as opposed to  mere {\sl bounds}) for the cross sections from first principles.

It is also known that, on the other hand, 
a behaviour of Regge type for the (imaginary part of the) scattering 
amplitude,
$$\imag F(s,t)\simeqsub_{s\to\infty}
\phiv(t)(\log s/\hat{s})^{\nu(t)}(s/\hat{s})^{\alpha_P(t)},
\equn{(1.2)}$$
where, for very small values of $t$,
$$\phiv(t)\simeq \ee^{bt},\quad\alpha_P(t)\simeq\alpha_P(0)+\alpha'_P(0)t,
$$
cannot hold for large values of the momentum transfer $|t|$.  
Here we expect the Brodsky--Farrar behaviour:\ref{6}
$$\imag F(s,t)\sim f(\cos\theta) s^{-p},\quad |t|\sim s,
\equn{(1.3)}$$
with $p$ is related to the number of constituents in particles $A,\;B$; 
for pion-pion scattering, $p=6$.
What is, however, not known is how the transition between the
 regimes described by (1.2) and (1.3) takes place.

In the present note we investigate the consequences of two possible assumptions 
that will allow us to make predictions for the high energy 
cross sections. 
First, we make what appears   a reasonable assumption, that we will call 
{\sl extended Regge} behaviour, suggested by   
Regge theory; this is given by 
Eq.~(1.2), assuming smooth behaviour of 
the trajectory $\alpha_P(t)$ and that 
(1.2) is valid up to $|t|\sim $ a few $\gev^2$; see below for 
precise details.
 This would allow us to refine the Froissart bound to the bounds
$$\dfrac{{\rm Const.}}{\log s/s_0}\lsim_{s\to\infty}
\sigma_{\pi\pi}(s)\lsim_{s\to\infty}({\rm Const.})\log s/s_0\quad\hbox{[Extended Regge]}
\equn{(1.4)}$$
for the pion-pion scattering amplitudes.
If, moreover, we further assume that partial wave amplitudes at fixed angular momentum, 
$l$, are mostly inelastic at high energy
 (actually, it is enough to assume that {\sl one} wave is inelastic) 
then we could mprove the bound for the
 cross section, getting
$$\lim_{s\to\infty}\sigma_{\pi\pi}(s)/\log s= 0\quad\hbox{[Extended Regge].}
\equn{(1.5)}$$ 

This bounds may be translated, via factorization,\ref{7} 
and adding the subleading $\rho$ and $P'$ Regge poles 
(necessary at --relatively-- low energies) to  the behaviour of 
$\pi N$, $KN$ and $NN$ cross sections at  high energies. 
If we do this, it turns out that the growth allowed 
by (1.4) is not very compatible with experimental 
cross sections at accessible energies, even if saturated,
 as it gives a  largish $\chidof\simeq1.3$. 
The predicted cross section  at the LHC would be
$$\sigma_{pp}(s)=((14\,{\rm TeV})^2)\sim95.5\pm4\;{\rm mb}\quad\hbox{[Extended Regge]}
\equn{(1.6)}$$
where the  error is only statistical (from fit to data).
This result is   difficult to believe; although compatible within errors, 
the number in (1.6) is clearly below  experiment\ref{5} already at 6~TeV.
We will discuss in \sect~6 the reasons for the failure of (1.5), essentially due to 
 failure of the Regge behaviour when $t$ is not near zero, as   
one must have functions  $\alpha_s(t)$, $\phiv(t)$   very different from what one expects in 
standard Regge theory. 

Then we consider a second possibility, which is that the bound (1.1b) 
is {\sl saturated}; we will give reasons that makes this saturation plausible. 
In this case, the fits to $NN$, $\pi N$ data improve clearly, and the prediction 
for the LHC cross section is
$$\sigma_{pp}(s)=((14\,{\rm TeV})^2)=108\pm4\pm4\;{\rm mb}\quad\hbox{[Saturated bound]}.
\equn{(1.7)}$$
Here the first error is statistical (from fit to data) and the second is the estimated 
theoretical error. 
The LHC data should be able to differentiate unambiguosly between this and 
the result from the extended Regge hypothesis, Eq.~(1.6).

\booksection{2. High energy cross sections with extended Regge behaviour}

\noindent
We will consider the $\pi^0\pi^+$ scattering amplitude, to avoid inessential 
complications associated with spin, isospin or identity of particles. 
Because of unitarity, we may write the scattering amplitude as
$$F_{\pi^0\pi^+}(s,t)=\sum_l(2l+1)P_l(\cos\theta) f_l(s),\quad
f_l(s)=\dfrac{2s^{1/2}}{\pi k}\dfrac{\eta_l(s)\ee^{2\ii\delta_l(s)}-1}{2\ii};\quad
\cos\theta=1+\dfrac{2t}{s-4\mu^2}.
\equn{(2.1)}$$
The elasticity parameter  is such that $0\leq \eta_l\leq1$. 
The new ingredient we use to get the bounds is that, in QCD, we have the Jackson--Farrar\ref{6} 
behaviour for the form factor of the pion, $F_\pi$,
$$F_\pi(s)\simeqsub_{s\to\infty}\dfrac{12\pi f^2_\pi\alpha_s(|s|)}{-s},\quad C_F=\tfrac{4}{3}.
\equn{(2.2)}$$
If we call $\delta_\pi(s)$ to the phase of $F_\pi(s)$, 
this implies that
$$\delta_\pi(s)\simeqsub_{s\to\infty}\pi
\left\{1+\dfrac{1}{\log s/\lambdav^2}\right\}.
\equn{(2.3)}$$
On the other hand, the Fermi--Watson final state theorem implies that, if the 
inelasticity is negligible, $\delta_\pi$ and the P wave phase,
$\delta_1$ are equal:
$$\delta_\pi\simeq\delta_1,\;{\rm if}\;\eta_1\simeq1.$$

Now comes the {\sl extended Regge} assumption: we assume the behaviour
$$\imag F_{\pi^0\pi^+}(s,t)\simeqsub_{s\to\infty}\tfrac{1}{3}
\phiv(t)\,(\log s/\hat{s})^{\nu(t)}(s/\hat{s})^{\alpha_P(t)}+
\hbox{(Const.)}. 
\equn{(2.4)}$$
The factor $\ffrac{1}{3}$ is a Clebsch--Gordan coefficient, that 
we separate off $\phiv$ for convenience.
We take $\hat{s}=1\,\gev^2$; the results, of course, are independent 
of this choice.
This behaviour is suggested by Regge theory. 
From general conditions it follows  that the functions $\phiv(t)$, 
$\nu(t)$ and $\alpha_P(t)$ must be analytic functions of $t$ in 
the Martin--Lehmannn ellipse. 
Moreover, from the positivity properties of $\imag F_{\pi^0\pi^+}(s,t)$ 
we expect $\alpha_P(t)$, $\nu(t)$, $\phiv(t)$ and all 
their derivatives to be be positive at $t=0$. 

We assume that $\alpha_P(t)$ is monotonously decreasing as $t$ becomes
 more and more negative, and that this happens for all values of $t$ 
provided $|t|\ll s$: this is the ``extended" hypothesis. 
In fact, it is sufficient to demand that this decrease occurs for  
 values  $|t|<|\tau_0|$, where $\tau_0$ is such that the integral 
$$\int_{-\infty}^{-\tau_0}\imag F_{\pi^0\pi^+}(s,t)
$$
is negligible, at large $s$. 
This hypothesis is, of course, verified if one had
$$\phiv(t)\simeq \ee^{bt},\quad\alpha_P(t)\simeq\alpha_P(0)+\alpha'_P(0)t,
$$
up to the value $t_0$ such that $\alpha_P(t_0)=0.$ 
From standard Regge fits, one expects $|t_0|\sim5\,\gev^2$.

\smallskip
\noindent{\sl Bounds}.\qquad
Integrating the imaginary part of (2.4) with $\tfrac{1}{2}\cos\theta$,
 we get the 
equality for the high energy 
P wave
$$\dfrac{4}{\pi}\,\dfrac{1-\eta_1\cos2\delta_1}{2}\simeqsub_{s\to\infty}\,
\dfrac{1}{2s}\int^0_{-\infty}\dd t\,
\phiv(t)(\log s/\hat{s})^{\nu(t)}(s/\hat{s})^{\alpha_P(t)}
\simeqsub_{s\to\infty}\,
\dfrac{1}{3\alpha'_P(0)}\phiv(0)\,
(\log s/\hat{s})^{\nu(0)-1}(s/\hat{s})^{\alpha_P(0)-1}.
\equn{(2.5)}$$
We have, in Eq.~(2.5), integrated with the formula (2.4) for all values of $t$. 
The fact that we can neglect the integral for large, negative values of $t$ is 
a consequence of the {\sl extended} Regge assumption: 
with it, the integral from any fixed $-\tau_0$ to $-\infty$ becomes negligible 
compared to the rest. 

Because the l.h.s in (2.5) is bounded, it follows that the r.h.s.
 must also be bounded and hence 
one must have $\alpha_P(0)\leq1$, $\nu(0)\leq1$ and, 
since
$$\sigma_{\pi^0\pi^+}(s)\sim s^{-1}\imag F_{\pi^0\pi^+}(s,0),$$ 
we get a first improvement of the Froissart bound:
$$\sigma_{\pi^0\pi^+}(s)\lsim_{s\to\infty}({\rm Const.})\log s/\hat{s}.
\equn{(2.6a)}$$
But we have more: if one had $\nu(0)<1$, 
then the r.h.s of (2.5) would tend to zero. So, the l.h.s. would also vanish 
which is only possible if $\eta_1=1$. 
In this case, the Fermi--Watson theorem applies 
and (2.3) gives
$$\dfrac{2}{\pi}\,\left\{1-\eta_1+2\pi^2
\left(\dfrac{1}{\log s/\lambdav^2}\right)^2\right\}\simeqsub_{s\to\infty}\,
\dfrac{1}{3\alpha'_P(0)}\,
\phiv(0)(\log s/\hat{s})^{\nu(0)-1}(s/\hat{s})^{\alpha_P(0)-1}.
$$
Because $1-\eta_1$ is positive, this is only possible if $\alpha_P(0)\geq1$
 and $\nu(0)-1\geq-2$
and we get the lower bound
$$
\sigma_{\pi^0\pi^+}(s)\gsim_{s\to\infty}\dfrac{{\rm Const.}}{\log s/\hat{s}},
\equn{(2.6b)}$$
which completes the lower and upper bound announced in (1.2). 

In fact, if we assumed that the partial wave amplitudes are mostly inelastic at 
high energy (actually, it is enough to assume that only one wave is inelastic) 
it follows tat the upper bound (2.6a) cannot be saturated as one must have 
$$\lim_{s\to\infty}\sigma_{\pi^0\pi^+}(s)/\log s=0.$$ 

\booksection{3. Saturated Froissart-like bound}

\noindent
We start by a brief derivation of the bound (1.1b). 
We write the Froissart--Gribov representation for the D wave in 
$\pi^+\pi^-\to\pi^0\pi^0$ scattering, $f_2(t)$, for $0<t\leq 4m^2_\pi$:
$$f_2(t)=\dfrac{1}{k_t^2}\dfrac{1}{\pi}\int_{4\mu^2}^\infty\dd s\,
\imag F_{\pi^0\pi^+}(s,t)Q_2\left(\dfrac{s}{2k_t^2}+1\right),
\qquad k_t=\dfrac{\sqrt{t-4\mu^2}}{2}.
\equn{(3.1)}$$
Because of elastic unitarity, it follows\ref{4,8} that the quantity 
$h(t)=f_2(t)/k_t^4$ and its two first derivatives at $t=4m^2_\pi$ 
are finite. 
For the second derivative, this implies a sum rule of the form
$$\int_{4\mu^2}^\infty\dd s\,
\dfrac{\partial^2\imag F_{\pi^0\pi^+}(s,t)/\partial t^2|_{t=4m^2_\pi}}{s^3}=C,
\equn{(3.2)}$$
and the constant $C$ can be expressed in terms of the
 scattering length and the two first effective 
radii of the D wave for $\pi^+\pi^-\to\pi^0\pi^0$ scattering. 
Because the derivatives of the Legendre polynomials 
$P_l(cos \theta)$ are positive for $\cos\theta\geq1$ ($t\geq0$), 
and grow rapidly with $l$, we get a bound for $\imag F_{\pi^0\pi^+}$. 
This comes about as follows:
the convergence of partial wave expansion,
$$\imag F_{\pi^0\pi^+}(s,t)=\sum_l(2l+1)P_l(\cos\theta) \imag f_l(s),$$
together with the fact that the $\imag f_l(s)$ are positive and bounded by 
$s^{1/2}/k\pi$, will allow us to  
 translate (3.2)  into the bound (1.1a) for the cross section. 
That  (3.2) is finite implies that, for large,  $s$,
$$\left.\dfrac{\partial^2\imag F_{\pi^0\pi^+}(s,t)}{\partial t^2}\right|_{t=4m^2_\pi}
=\dfrac{4}{(s-4m^2_\pi)^2}\sum_{l=0}^\infty(2l+1)P''_l(\cos\theta) \imag f_l(s)<
({\rm Const.})s^2.
$$
We can now use this to bound the sum $\sum_{l=l_0}^\infty$ 
in the expression for $\imag F$, and the unitarity bound 
$\imag f_l\leq 2s^{1/2}/\pi k$ for the piece 
$\sum_{l=0}^{l_0}$. 
Optimizing $l_0$ (which one takes $l_0\sim s^{1/2}$) 
produces 
 the bound (1.1b);
the details may be found in ref.~4.
This bound improves the standard  Froissart bound in (1.1a), that can be obtained from 
 (3.1), because ${P_l}''(cos \theta)\gg {P_l}'(cos \theta)\gg {P_l}(cos \theta)$, 
for $t\geq0$, large $l$.
This bound (1.1b) is optimal, in the sense that, if we take a further derivative, 
the corresponding integral,
$$\int_{4\mu^2}^\infty\dd s\,
\dfrac{\partial^3\imag F_{\pi^0\pi^+}(s,t)/\partial t^3|_{t=4m^2_\pi}}{s^3},
$$
necessarily diverges.

This suggests a behaviour like (1.1b), that is,
$$\sigma_{\pi^0\pi^+}(s)\simeqsub_{s\to\infty}({\rm Const.})\times
\log^2\dfrac{s}{s_1\log^{\frac{7}{2}}s/s_2}.
\equn{(3.3)}$$

\booksection{4. Phenomenology}

\noindent
In order to compare our result with very high energy $\pi N$, $KN$ and 
(especially) $NN$ 
cross sections, for which we have good data, we may use factorization to write
$$\eqalign{
\sigma_{\pi\pi}^{(I_t=0)}\simeqsub_{s\;{\rm large}}&
\;\dfrac{4\pi^2}{\lambda^{1/2}(s,m_{\pi}^2,m_{\pi}^2)}\Big[P(s,0)+P'(s,0)\big],\cr
\dfrac{\sigma_{pp}+\sigma_{\bar{p}p}}{2}
\simeqsub_{s\;{\rm large}}&\;\dfrac{4\pi^2}{\lambda^{1/2}(s,m_p^2,m_p^2)}
\,\tfrac{1}{2}f_{N/\pi}^2\Big[P(s,0)+(1+\epsilon)P'(s,0)\Big],\cr
\sigma_{\pi^\pm p}
\simeqsub_{s\;{\rm large}}&\;\dfrac{4\pi^2}{\lambda^{1/2}(s,m^2_\pi,m_p^2)}f_{N/\pi}
\left\{\dfrac{1}{\sqrt{6}}\Big[P(s,0)+P'(s,0)\Big]\mp
\tfrac{1}{2}\bar{\rho}(s,0)\right\},\cr
\left.\dfrac{\dd\sigma(\pi^- p\to\pi^0 n)}{\dd t}\right|_{t=0}
\simeqsub_{s\;{\rm large}}&\;f_{N/\pi}^2\,
\dfrac{1-\cos\pi\alpha_\rho}{\sin^2\pi\alpha_\rho}\,
\dfrac{\pi^3}{\lambda(s,m^2_\pi,m_p^2)}\,\left|\bar{\rho}(s,0)\right|^2\cr
\sigma_{K^+ p}+\sigma_{K^- p}
\simeqsub_{s\;{\rm large}}&\;\dfrac{4\pi^2}{\lambda^{1/2}(s,m^2_K,m_p^2)}f_{N/\pi}f_{K/\pi}
\Big[P(s,0)+rP'(s,0)\Big].
\cr}
\equn{(4.1a)}$$ Moreover, 
$$\bar{\rho}(s,0)=\bar{\beta}_\rho\,
(s/\hat{s})^{\alpha_\rho(0)},\quad 
P'(s,0)=\beta_{P'}(s/\hat{s})^{\alpha_{P'}(0)}.
\equn{(4.1b)}$$

The quantities  
$\alpha_\rho(0)$, $\bar{\beta}_\rho$  
have been determined with precision to be\ref{9}
$$
\bar{\beta}_\rho=\alpha_{P'}=0.39\pm0.02,\quad
\alpha_\rho(0)=0.52\pm0.03.
\equn{(4.1b)}$$
One also has $\epsilon=0.24$ and $r$ is very  small.

\noindent{\sl The extended Regge case.}
\qquad We have shown before that the assumption of an extended Regge behaviour 
leads to a growth of cross sections less than $\log s$. 
This is not sufficient to reproduce the rise of the cross sections
 for hadronic processes  observed to occur in the multi-TeV region.
We may verify this if, 
in the extended Regge case, we take the most favourable 
situation in which the bound (2.6a) is saturated 
and thus write
$$
P(s,0)=\left\{a\log\dfrac{s}{\hat{s}}+\widetilde{\beta}_P\right\}
s,\quad \hat{s}\equiv1.
\equn{(4.2)}$$ 
With this, we  fit data for $\pi\pi$, $\pi^\pm p$, $K^+p+K^-p$,
 and $pp+\bar{p}p$   cross
sections\fnote{For details on the choice of 
data, errors, etc., see ref.~9.}
from a kinetic energy in the c.m. of one GeV  up
to the highest energies attained, 
30~TeV in cosmic ray experiments.\ref{10}  
We  find
$$a=0.413\;{\rm mb},\quad \widetilde{\beta}_P=-1.36,\quad 
f_{N/\pi}=1.320,\quad f_{K/\pi}=0.831.
\equn{(4.3)}$$
The fit is not very good, as one has $\chidof=1.29$. 
The value of the corresponding cross section at the LHC is as reported in (1.6), 
certainly too low.

\noindent{\sl The saturated Froissart bound case.}
\qquad
The {\sl saturated bound} hypothesis fares  better. 
We  write now,
$$P(s,0)=\left\{A\,\log^2\dfrac{s}{s_1\log^{\frac{7}{2}}s/s_2}+\widetilde{\beta}_P\right\}s.
\equn{(4.4)}$$
The details of the fit may be found in ref.~9. We choose the 
fit called ``fit C" there and have
$$\eqalign{f_{N/\pi}=&\,1.359\pm0.004,\quad
\widetilde{\beta}_P=2.32\pm0.04,\cr
 A=&\,0.033\pm0.001,\quad s_1=0.01\,\gev^2,
\quad s_2=0.15\pm0.05\,\gev^2.\cr
}
\equn{(4.5)}$$
The fit has chi-squared of $\simeq1.2$ which, for almost 500 
experimental points, is clearly better than that 
with the 
extended Regge hypothesis (particularly if we realize 
that a \chidof\ of 1.15 would be obtained if 
relaxing the extended factorization hypothesis, and a 
\chidof=1 would follow if excluding $\pi^+p$ data for $s^{1/2}<3\,\gev$; 
cf.~ref.~9). 
The fit is depicted, for some of the processes, in \fig~1.

\topinsert{
\setbox3=\vbox{\hsize16truecm{\epsfxsize 14.0truecm\epsfbox{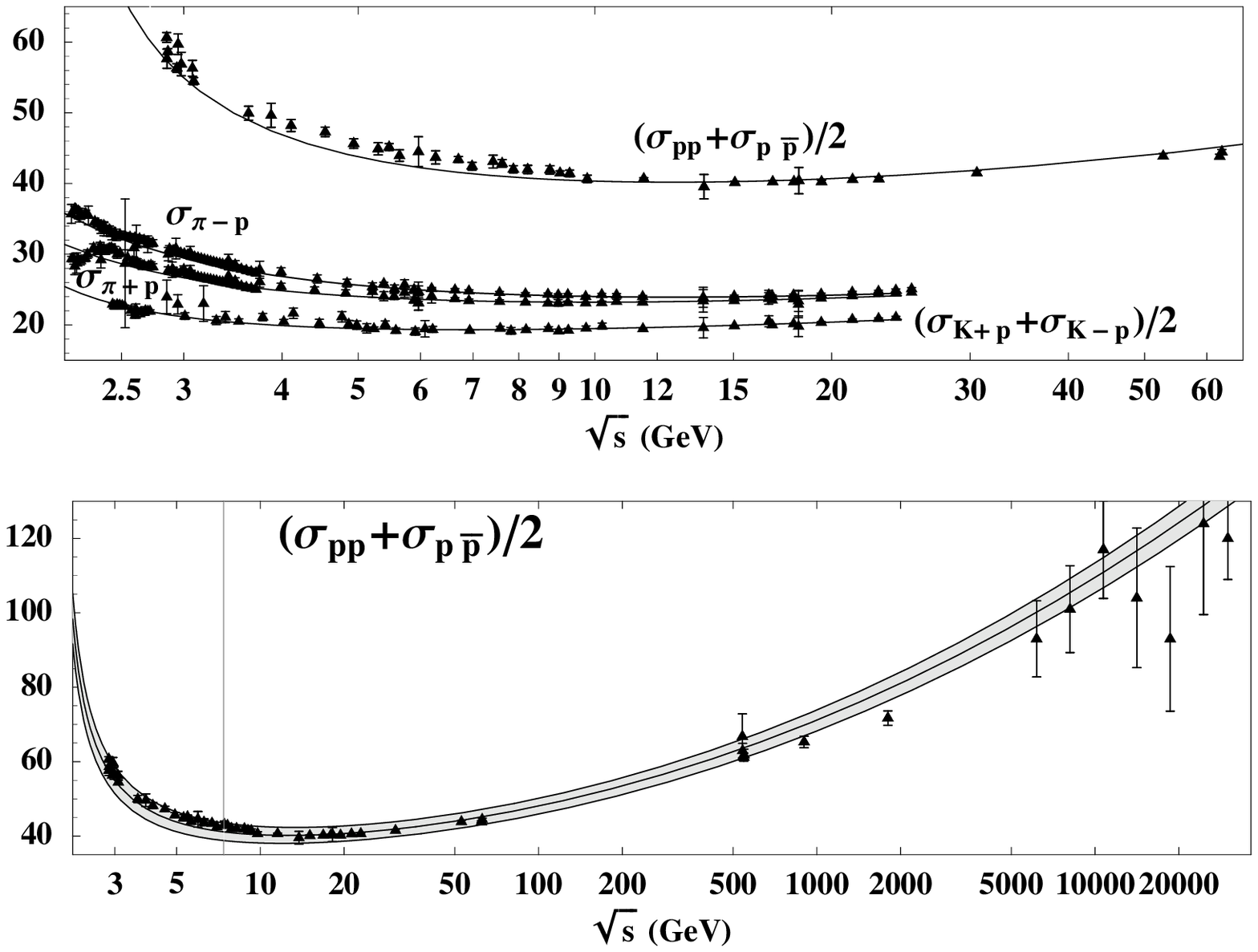}}}
\centerline{{\box3}\hfil}
\bigskip
\setbox5=\vbox{\hsize 14.5truecm\captiontype\figurasc{Figure 1. }
 {The total cross section
 $(\sigma_{\bar{p}p}+\sigma_{pp})/2$. 
Black dots, triangles and squares: experimental points. 
For energies above 30 \gev, we have given the experimental values of 
 $(\sigma_{\bar{p}p}+\sigma_{pp})/2$ as if they equaled $\sigma_{\bar{p}p}$
 or $\sigma_{pp}$.
Continuous lines: fits with the saturated Froissart bound 
hypothesis.}}
\centerline{\box5}
\bigskip
}\endinsert

\booksection{5. Discussion}

\noindent
The fact that it is impossible to get a good fit with the 
parametrization (4.2) in the whole 
energy range $1\,\gev\lsim E_{\rm kin}\lsim30\,{\rm TeV}$, 
while (4.4) produces a clearly better  
fit in the same energy region, 
suggests that  it is the saturated Froissart bound 
hypothesis, and not the 
extended Regge  one, that makes the ultra high energy 
behaviour of cross sections  compatible with 
 what one finds at (relatively) lower energies.
That the extended Regge behaviour fails means that the 
behaviour (2.4) must only hold for 
values of $t$ near zero.
In fact, it is not difficult to realize 
that the behaviour (1.1b) is only compatible with 
the saturated Froissart bound, Eq.~(3.3), under the following conditions:   the   
function $\alpha_s(t)$ in (1.2) must flatten out before vanishing. 
Moreover, if we call $t_0$ to the point 
where $\alpha_s(t)$ first vanishes, the residue function $\phiv(t)$ must change sign 
at relatively small values of $t$ and continue negative until $t\sim t_0$ 
and, furthermore,  
 $t_0$ must be of the order of $s$, for large $s$. 
This is so because one must cancel, to a relative precision 
$O(1/\log s)$, the 
integral
$$\int^0_{-t_1}\dd t\,P_l(\cos\theta)\imag F_{\pi^0\pi^+}(s,t)\sim \log s$$
with the remainder,
$$\int^{-t_1}_{4m^2_\pi-s}\dd t\,P_l(\cos\theta)\imag F_{\pi^0\pi^+}(s,t),$$
where $t_1$ is the first zero of $\phiv(t)$.
Now, since $\alpha_P(t)$ is assumed to be independent of $s$, it follows that 
one should have $\alpha_P(t)>0$ for all $t$.
In particular, it follows that the transition from the 
Regge behaviour to the Brodsky--Farrar one would be very rough, 
involving violent oscillations of the scattering amplitude.
Indeed, to get from a behaviour $\imag F_{\pi^0\pi^+}\sim\phiv s^0$ to 
 $\imag F_{\pi^0\pi^+}\sim s^{-6}$ one should have very strong oscillations of 
$\phiv(t)$ 
which would average to zero when $|t|\to\infty$.
 
This is very unlikely; what probably happens is that 
 the classical Regge-type expression,\ref{11}
$$P(s,t)=a(\log^\nu s/\hat{s})\alpha_P(t)\,\dfrac{1+\alpha_P(t)}{2}\,
\ee^{bt}(s/\hat{s})^{\alpha_P(t)},\quad
\alpha_s(t)\simeq1+\alpha'_P(0)t,$$
 fails well before the point $t_0$ where $\alpha_P(t_0)=0$, 
i.e., well below $|t|\sim5\,\gev^2$.

\vfill\eject
\noindent{\sl Acknowledgments}

\noindent
I very am grateful to J.~R.~Pel\'aez who collaborated in parts of this work.

\noindent{\bf References}

\item{1 }{ Froissart, M. {\sl Phys. Rev.} {\bf 123}, 1053 (1961).}
\item{2 }{ Martin, A. {\sl Nuovo Cimento}, {\bf 42}, 930 (1066);
 Epstein,~H., Glaser,~V., and Martin,~A. 
{\sl Commun. Math. Phys.}, {\bf 13}, 257 (1969).}
\item{3 }{ L.~\L ukaszuk and A.~Martin {\sl Nuovo Cimento}, {\bf 52A}, 122 (1067);
 F. J.~Yndur\'ain, {\sl Phys. Letters} {\bf 31B}, 358, 620 (1970); D.~Atkinson, in 
``Strong Interaction Physics", G.~H\"oller (Ed.) Springer Tracts {\bf 57}, Berlin, 1971.}
\item{4 }{F. J.~Yndur\'ain {\sl Phys. Letters}, {\bf 42B}, 571 (1972).}
\item{5 }{Cf. the compilations of the COMPAS group, IHEP, Protvino, Russia.
 [See the Particle Data Tables: Hagiwara, K., et al., {\sl Phys. Rev.} 
{\bf D66}, 010001 (2002)].}
\item{6 }{Brodsky, S. J., and Farrar, G., {\sl Phys. Rev. Lett.}, {\bf 31}, 1153  (1973);  
Farrar, G., and Jackson, D. R., {\sl Phys. Rev. Lett.}, {\bf 43}, 246 (1979).}
\item{7 }{Gell-Mann, M.  {\sl Phys. Rev. Letters}, {\bf 8}, 263, (1962); 
Gribov, V. N., and Pomeranchuk, I. Ya.  {\sl Phys. Rev. Letters}, {\bf 8}, 343,  (1962).
For QCD, cf. Gribov, V. N., and  Lipatov, L. N. {\sl Sov. J. Nucl. Phys.} {\bf 15}, 438 and 675 
(1972); 
Kuraev, E. A., Lipatov, L. N., and Fadin, V. S.
 {\sl Sov. Phys. JETP} {\bf 44}, 443 (1976); 
Dokshitzer,~Yu.~L.  {\sl Sov. Phys. JETP} {\bf 46,} 641 (1977); 
Balitskii, Ya. Ya., and  Lipatov, L. N.  {\sl  Sov. J. Nucl. Phys.} {\bf 28}, 822 (1978).}
\item{8 }{F. P. Palou, and  F. J. Yndur\'ain, {\sl Nuovo Cimento}, {\bf 19A}, 245, 
 (1974).}
\item{9 }{Pel\'aez, J. R.,  and Yndur\'ain, F. J., hep-ph/0312187, in 
press in Phys.~Rev.~D.}
\item{10 }{Cosmic ray results: R. M. Baltrusaitis, et al.  {\sl Phys. Rev. Lett.} {\bf 52}, 1380 (1984); 
M.~Honda, et al. {\sl Phys. Rev. Lett.} {\bf 70}, 525 (1993).}
\item{11 }{Rarita, W., et al., {\sl Phys. Rev.} {\bf 165}, 1615, (1968).}

\bye